\documentclass[10pt,twocolumn,letterpaper]{article}

\usepackage{cvpr}              % To produce the CAMERA-READY version
\usepackage{listings}
\definecolor{cvprblue}{rgb}{0.21,0.49,0.74}
\usepackage[pagebackref,breaklinks,colorlinks,allcolors=cvprblue]{hyperref}
\usepackage{tabularx}
\usepackage{booktabs}
\usepackage{caption}
\usepackage{comment}

\def\confName{CVPR}
\def\confYear{2025}

\title{
\textbf{Forecasting Labor Markets with LSTNet: A Multi-Scale Deep Learning Approach} \\ [-2.3ex]
} 

\author{
\textbf{Omar Abbouchi}, \quad 
\textbf{Sofia Davila}, \quad 
\textbf{Meena Al Hasani}, \\
\textbf{Jessica Le}, \quad 
\textbf{Adam Nelson-Archer}, \quad 
\textbf{Aleia Sen} \\ [1.2ex]
\normalfont University of Houston Department of Computer Science
\vspace{-0.5em}
}

\begin{document}
\maketitle 
\begin{abstract}
    We present a deep learning approach for forecasting short-term employment changes and assessing long-term industry health using labor market data from the U.S. Bureau of Labor Statistics. Our system leverages a Long- and Short-Term Time-series Network (LSTNet) to process multivariate time series data, including employment levels, wages, turnover rates, and job openings. The model outputs both 7-day employment forecasts and an interpretable Industry Employment Health Index (IEHI). Our approach outperforms baseline models across most sectors, particularly in stable industries, and demonstrates strong alignment between IEHI rankings and actual employment volatility. We discuss error patterns, sector-specific performance, and future directions for improving interpretability and generalization.
\end{abstract}

\section{Introduction}
\label{sec:Problem_Statement}

% Added a hook to our abstract
The labor market's complex and dynamic nature presents significant challenges for employment forecasting and analysis. Traditional approaches often fail to capture the nuanced patterns in employment data, particularly the interplay between short-term fluctuations and long-term trends across different industries. This limitation affects various stakeholders, from job seekers trying to make informed career decisions to policymakers developing labor market strategies.

We propose an approach to employment forecasting using a Long- and Short-term Time-series Network (LSTNet) architecture. Our system addresses two challenges: accurate short-term employment forecasting and comprehensive industry health assessment. The LSTNet model processes multivariate industry-level time series data from the Bureau of Labor Statistics, including unemployment rates, employment level, job turnover data, average hours worked per week, and average weekly earnings, to generate both employment predictions and an Industry Employment Health Index (IEHI).

The system's architecture combines convolutional neural networks and gated recurrent units to capture both immediate patterns and long-term dependencies in employment data. Skip connections handle periodic patterns, while an autoregressive component models short-term trends. This multi-scale approach enables the model to generate accurate 7-day forecasts while maintaining interpretability through the IEHI score.

Our implementation demonstrates promising results in processing weekly employment data from 2006 to the present. From our initial work, we have expanded the dataset to include additional economic indicators and industry-specific metrics, further enhancing the model's predictive capabilities and the IEHI's comprehensiveness.

The system's inputs include the following: 
 \begin{itemize}
     \item Historical employment rates and unemployment data for the industry 
     \item Job vacancy and hiring data, e.g. open positions, hires, and separations
     \item Employee-level data, e.g. average weekly hours and salary
 \end{itemize}
 The system outputs an employment change forecast and an IEHI value for each industry. 

\begin{table*}[t]
    \centering
    \caption{An example of weekly data processed. This shows sector-level employment data from December 2024, including workforce size, wages, hours, and labor dynamics.}
    \label{tab:employment}
    \footnotesize
    \begin{tabularx}{\textwidth}{l|r|r|r|r|r|r|r}
    \toprule
    Sector & Employees (K) & Avg Hours & Avg Earnings (\$) & Unemp. Rate & Openings (K) & Hires (K) & Separations (K) \\
    \midrule
    Construction & 8289.0 & 38.8 & 38.94 & 5.2\% & 205.0 & 327.0 & 268.0 \\
    Education and Health Services & 26931.0 & 32.9 & 35.01 & 2.7\% & 1518.0 & 625.0 & 677.0 \\
    Financial Activities & 9206.0 & 37.7 & 46.37 & 2.1\% & 390.0 & 164.0 & 168.0 \\
    Information & 2944.0 & 36.9 & 51.04 & 3.9\% & 105.0 & 50.0 & 62.0 \\
    Leisure and Hospitality & 16979.0 & 25.5 & 22.40 & 5.4\% & 998.0 & 675.0 & 804.0 \\
    Manufacturing & 12760.0 & 40.1 & 34.54 & 3.5\% & 398.0 & 204.0 & 249.0 \\
    Natural Resources & 624.0 & 44.3 & 39.95 & 5.4\% & 21.0 & 15.0 & 23.0 \\
    Other Services & 6002.0 & 32.0 & 32.37 & 3.8\% & 248.0 & 188.0 & 198.0 \\
    Professional Services & 22614.0 & 36.3 & 43.33 & 3.7\% & 1276.0 & 712.0 & 937.0 \\
    Transportation and Utilities & 29033.0 & 34.0 & 30.34 & 4.3\% & 988.0 & 913.0 & 1117.0 \\
    \bottomrule
    \end{tabularx}
\end{table*}
\section{Related Work}
\label{sec:Related_Work}
Recent work in labor market forecasting and time series modeling has demonstrated the effectiveness of deep learning approaches in predicting employment trends. One example is recent work by Yurtsever \cite{yurtsever2023unemployment}, who applied recurrent neural networks (LSTM-GRU) to forecast national unemployment rates in the US, the UK, France, and Italy. This method provides a benchmark for comparing classical models with state-of-the-art techniques.

Salinas et al.\cite{salinas2020deepar} introduced DeepAR, a global recurrent neural network capable of modeling numerous related time series simultaneously. This approach enables generalization across industries, particularly when data availability varies by sector.

Further advancements in forecasting have come from multivariate LSTM-based models, which can integrate external variables such as economic indicators, job postings, and turnover rates into predictive frameworks. Lai et al. \cite{lai2018modeling} developed LSTNet, a hybrid convolutional and recurrent neural network that captures short-term fluctuations and long-term dependencies in multivariate time series. Similarly, Ju and Liu \cite{ju2021multivariate} proposed ATT-LSTM, which enhances LSTM networks with attention mechanisms to effectively filter influences between variables.

% added new stuff below %

Recent developments in transformer-based models, such as the Temporal Fusion Transformer (TFT) introduced by Lim et al. \cite{lim2021temporal}, offer significant improvements in modeling complex interactions between time series variables. TFTs employ attention mechanisms to provide interpretability by quantifying the contribution of individual input features.  

Hybrid approaches that combine classical econometric methods with deep learning have also demonstrated promising results. Deep Vector Autoregression (Deep VAR), proposed by Altmeyer et al. \cite{agusti2021deep}, enhances traditional VAR models by capturing nonlinear economic interactions and regime shifts through the use of neural networks. Deep VAR similarly targets multivariate dependencies, capturing nonlinear interactions through deep learning extensions to classical models.

Major events in recent history have emphasized the need for accurate and adaptable forecasting models. Specifically, events like the COVID-19 pandemic have highlighted the need for predictive models. A comprehensive review by Alzubaidi et al.\cite{alzubaidi2022behavioral} analyzed 53 studies employing deep learning for COVID-19 forecasting. Additionally, Dairi et al.\cite{dairi2021comparative} conducted a comparative study of hybrid models in COVID-19 forecasting. Both reviews found that LSTM networks and their variants (e.g., BiLSTM, ConvLSTM) were the most widely used and generally outperformed traditional statistical models in capturing the complex dynamics of the pandemic.

To specifically address unemployment and economic forecasting, traditional approaches have often relied on econometric methods like ARIMA \cite{montgomery1998making} due to their interpretability and ease of use. However, contemporary research, such as a recent comparative analysis of forecasting models by Apte and Haribhakta\cite{apte2024advancing} increasingly highlights the limitations of these methods in capturing the nonlinear and dynamic nature of modern employment data.  Alternative models, such as N-HiTS and N-BEATS, have demonstrated superior performance in capturing complex patterns and nonlinear dynamics in time series data\cite{zhang2023time}.

Employment indices such as the Conference Board’s Employment Trends Index and JOLTS data \cite{barnichon2010building,stock2003forecasting} aggregate multiple labor market indicators into actionable metrics. Recent research has emphasized building composite indices that integrate predictive modeling with labor market indicators, offering more dynamic assessments compared to traditional indices. For example, Heise et al. \cite{heise2024wage} proposed the HPW Tightness Index, which aggregates labor market variables to better explain wage growth dynamics. Similarly, Ebadi \cite{ebadi2025rule} introduced the E-Rule, a composite recession indicator combining financial and labor market signals, while Kim et al. \cite{kim2022development} applied LSTM models to predict labor turnover patterns more accurately than autoregressive approaches.
\section{Dataset}
\label{sec:dataset}

Our system processes multivariate time series data from the Bureau of Labor Statistics (BLS)\cite{bls-bed} to forecast employment trends. The dataset consists of weekly employment indicators spanning from March 2006 to December 2024, showcased in Table~\ref{tab:employment}. For each of the ten industries utilized, the data instance contains the employment level, the average weekly hours, the average weekly earnings, the unemployment rate, the number of openings, the number of hires, and the number of separations. The BLS reports this data in monthly intervals, which have been interpolated to obtain weekly approximations for a higher level of granularity.

The BLS API organizes each series by the specific metric and industry for which the data is reported. The dataset was built by querying the API with a JSON payload for each series, containing the series identification code, the start year, the end year, and the API key. Each observation is indexed by the weekly timestamp and the industry, as well as the corresponding values for each employment metric. The dataset has been preprocessed to handle missing values and to ensure temporal consistency. 

The model processes this multivariate time series through a 30-day sliding window, allowing it to capture both immediate trends and longer-term patterns in the employment data. The output consists of:
\begin{itemize}
    \item A 7-day forecast of employment changes
    \item An Industry Employment Health Index (IEHI) score derived from the model's predictions and feature importance analysis
\end{itemize}

This dataset structure enables our LSTNet model to learn complex temporal dependencies and generate both short-term forecasts and industry health assessments. The weekly frequency of the data provides sufficient resolution for our forecasting objectives while maintaining data quality and consistency. 

\begin{figure*}[t]
 \centering
 \includegraphics[width=0.9\textwidth]{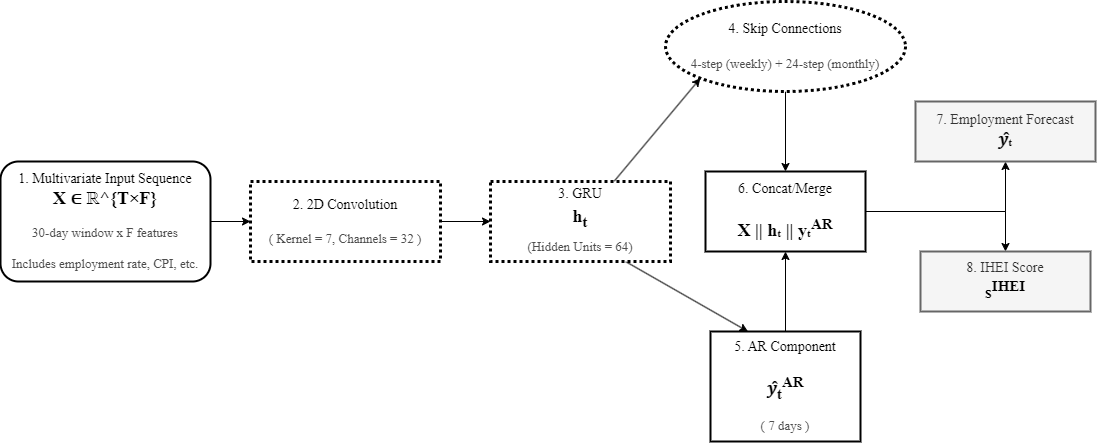}
 \caption{Overview of our LSTNet-based architecture for employment forecasting and industry health assessment.}
 \label{fig:model_architecture}
\end{figure*}

\section{Evaluation Metric}
\label{sec:Evaluation_Metric}

Accurate evaluation of time-series forecasts is essential for assessing model performance across industries with varying employment dynamics. Our evaluation framework is designed to capture both the magnitude of prediction error and the structural alignment of forecasts with actual trends. We employ three error-based metrics—Symmetric Mean Absolute Percentage Error (SMAPE), Root Mean Squared Error (RMSE), and Mean Absolute Error (MAE)—and one rank-based metric, Spearman’s rank correlation coefficient ($\rho$), to provide a comprehensive analysis.

\subsection{Symmetric Mean Absolute Percentage Error (SMAPE)}
\[
\text{SMAPE} = \frac{100}{n} \sum_{i=1}^{n} \frac{|y_i - \hat{y}_i|}{(|y_i| + |\hat{y}_i|)/2}
\]

SMAPE measures the average relative error in percentage terms \cite{makridakis2000m3} while addressing the asymmetry and instability inherent in the standard Mean Absolute Percentage Error (MAPE), particularly when actual values are near zero. As our forecasting task involves predicting employment *changes*, which often oscillate around zero due to normalization or smoothing, MAPE was found to yield disproportionately large error values that failed to reflect visual or structural accuracy. SMAPE, by contrast, bounds the error between 0\% and 200\%, ensuring comparability across industries with different levels of employment volatility. In our experiments, SMAPE provided the most interpretable measure of performance across domains, especially when comparing volatile sectors such as mining, manufacturing, and information technology.

\subsection{Root Mean Squared Error (RMSE)}
\[
\text{RMSE} = \sqrt{ \frac{1}{n} \sum_{i=1}^{n} (y_i - \hat{y}_i)^2 }
\]

RMSE quantifies the magnitude of the forecast error and is especially sensitive to large deviations\cite{hyndman2018forecasting}. This makes it useful for penalizing models that occasionally produce large prediction failures, even if average performance is otherwise reasonable. RMSE retains the original units of the dependent variable, allowing for interpretation in absolute employment change terms. We include RMSE to complement SMAPE and capture the practical cost of over- or under-estimating job market shifts.

\subsection{Mean Absolute Error (MAE)}
\[
\text{MAE} = \frac{1}{n} \sum_{i=1}^{n} |y_i - \hat{y}_i|
\]

MAE provides a linear measure of average error that is more robust to outliers than RMSE. While it lacks the percentage-based normalization of SMAPE, MAE is particularly useful in understanding model consistency. In our analysis, sectors with high MAE but moderate SMAPE—such as Leisure and Hospitality—highlight cases where the model tracks the general trend well but suffers from absolute misalignment due to seasonality or structural breaks.

\subsection{Spearman's Rank Correlation Coefficient ($\rho$)}
To assess the ability of our model to rank industries by employment growth potential—critical for our hiring index—we employ Spearman’s $\rho$, defined as the Pearson correlation of the rank values of the variables. Unlike standard correlation coefficients, Spearman’s $\rho$ evaluates monotonic relationships and is therefore more robust to non-linearities in the underlying data.

Formally:
\[
\rho = 1 - \frac{6 \sum d_i^2}{n(n^2 - 1)}
\]
where $d_i$ is the difference in rank between predicted and actual values.

This metric allows us to evaluate how well our model preserves ordinal relationships between industries, even when precise forecasts may deviate. A high $\rho$ value indicates that the model effectively identifies which sectors are improving or declining relative to others, aligning with the decision-making goals of our hiring index.

\subsection{Metric Selection}

Although MAPE is commonly used in economic forecasting \cite{hyndman2006another, armstrong1985long,flores1986pragmatic}, it was found to be unsuitable for our application due to the normalized nature of employment change data and its high sensitivity to near-zero values\cite{goodwin1999forecasting}. As such, we excluded MAPE from final evaluation in favor of SMAPE, which more accurately reflects model performance across sectors, particularly those with low base employment levels or high relative volatility. Our decision was further supported by observed discrepancies in model error interpretation, where models with visually close forecasts yielded MAPE values exceeding 100\%, whereas SMAPE remained within a consistent and interpretable range.

\begin{figure*}[t]
 \centering
 \includegraphics[width=1.0\textwidth]{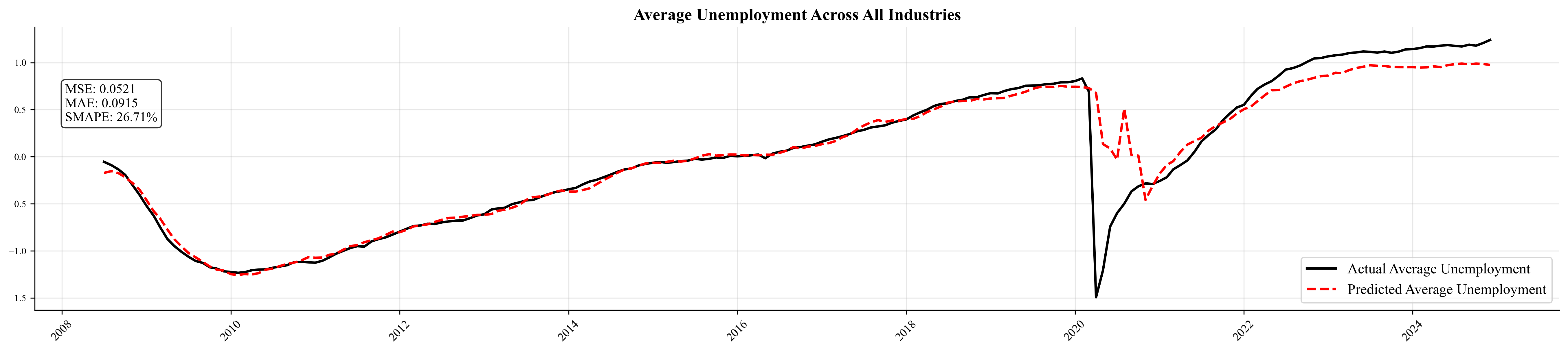}
 \caption{Average Change in Employment Across All Industries, 2008 through 2024 - Actual vs. Predicted}
 \label{fig:overall_employment}
\end{figure*}

\begin{figure}[t]
 \centering
 \includegraphics[width=1\linewidth]{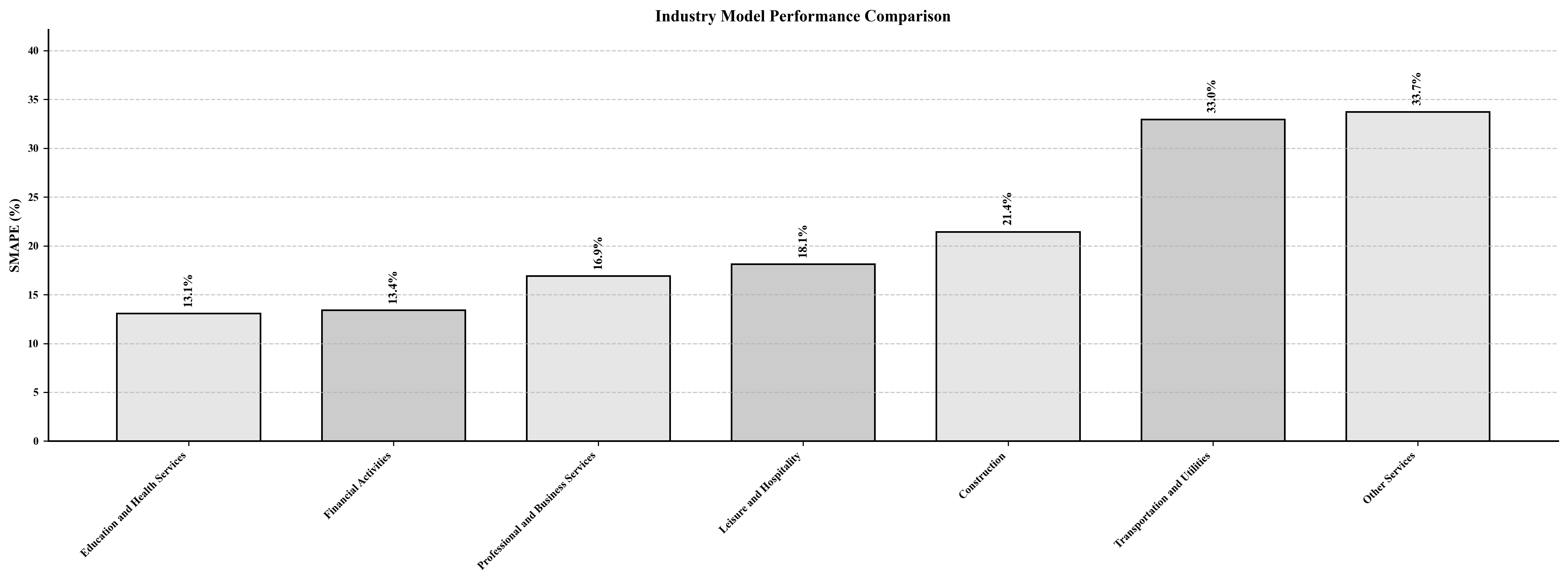}
 \caption{SMAPE (\%) by Industry. Ordered from Least to Greatest, "Education and Health Services", "Financial Activities", "Professional Services", "Leisure and Hospitality", "Construction", "Transportation and Utilities", "Other Services"}
 \label{fig:model_performance}
\end{figure}
\section{Baseline and Oracle Evaluation}
\label{sec:Baseline_Oracle}

To contextualize the performance of our LSTNet-based forecasting system, we evaluate two comparative models: a naïve persistence baseline and a theoretical oracle. These benchmarks define the lower and upper bounds for expected model performance across diverse labor market segments.

\subsection{Persistence Baseline}

The persistence model is a simple heuristic that assumes employment metrics remain unchanged over the prediction horizon. Given an input sequence, the forecast for the next seven days replicates the final observed value from the input. This model captures the inertia typical in stable economic conditions but fails to adapt to dynamic shifts. Its implementation is computationally trivial and requires no training.

Formally, if \( y_t \) is the most recent observation, then for forecast horizon \( h \), the predicted sequence \( \hat{y}_{t+1}, \hat{y}_{t+2}, \dots, \hat{y}_{t+h} \) is defined as:
\[
\hat{y}_{t+i} = y_t, \quad \text{for } i = 1, \dots, h
\]

\subsection{Oracle Approximation}

To estimate an aspirational upper bound, we implemented an oracle model that uses future ground-truth data during inference. Specifically, it predicts the mean of the actual values over the forecast horizon. Although this violates causality and is not deployable in practice, it helps to quantify the intrinsic unpredictability of different industry sectors under ideal conditions.

The oracle forecast for a horizon of length \( h \) is:
\[
\hat{y}_t^{oracle} = \frac{1}{h} \sum_{i=1}^{h} y_{t+i}
\]

\subsection{Evaluation and Results}
Both models were implemented within a unified evaluation pipeline using identical preprocessing, sequence generation, and metric computation steps as the primary LSTNet model. For each industry, we report two standard regression metrics—Root Mean Squared Error (RMSE) and Symmetric Mean Absolute Percentage Error (SMAPE)—evaluated on the employment time series.

Table~\ref{tab:baseline_performance} presents these results. As expected, the oracle model, which uses ground-truth future values, yields the lowest error values across all sectors. In contrast, the persistence model performs reasonably in more stable sectors (e.g., Education and Health Services), but its accuracy degrades sharply in volatile industries such as Information and Other Services. Notably, the persistence model's SMAPE values are highly inflated in domains with small denominators, illustrating the sensitivity of percentage-based metrics when actual values approach zero.

\begin{table}[htbp]
    \centering
    \scriptsize
    \caption{Baseline Model Performance by Industry}
    \label{tab:baseline_performance}
    \setlength{\tabcolsep}{5pt} % Reduce column spacing
        \begin{tabular}{l@{\hskip 4pt}c@{\hskip 4pt}c@{\hskip 7pt}c@{\hskip 4pt}c}
        \toprule
        & \multicolumn{2}{c}{Oracle Model} & \multicolumn{2}{c}{Persistence Model} \\
        \cmidrule(lr){2-3} \cmidrule(lr){4-5}
        Industry & RMSE & SMAPE (\%) & RMSE & SMAPE (\%) \\
        \midrule
        Construction & 0.104 & 12.43 & 1.01 & 170.16 \\
        Education & 0.088 & 9.67 & 0.87 & 148.50 \\
        Financial Activities & 0.05 & 6.92 & 1.06 & 196.62 \\
        Information & 0.173 & 23.60 & 0.90 & 140.29 \\
        Leisure/Hospitality & 0.380 & 12.66 & 1.00 & 166.15 \\
        Manufacturing & 0.157 & 15.54 & 0.66 & 120.41 \\
        Natural Resources & 0.091 & 20.08 & 1.06 & 197.63 \\
        Other Services & 0.455 & 15.06 & 1.05 & 176.54 \\
        Professional Services & 0.087 & 9.52 & 1.00 & 160.80 \\
        Transportation/Utilities & 0.184 & 12.12 & 1.06 & 183.71 \\
        \midrule
        Average & 0.177 & 13.76 & 0.97 & 166.08 \\
        \bottomrule
    \end{tabular}
\end{table}

\section{Main Approach}
\label{sec:mainApproach}

Our approach addresses the dual objective of short-term employment forecasting and long-term sector health assessment through a specialized Long- and Short-term Time-series Network (LSTNet) architecture. The model takes in multivariate time series data comprising historical employment trends, macroeconomic indicators (e.g., GDP, inflation, consumer sentiment), labor market dynamics (e.g., vacancy rates, applicant-to-job ratios), and static industry-specific features such as skill requirements and workforce retention rates~\cite{stock2003forecasting,barnichon2010building}. These inputs are used to generate weekly employment change predictions as well as an interpretable Industry Employment Health Index (IEHI), which reflects structural resilience and volatility.

The core of our LSTNet implementation consists of a multi-layer architecture designed to capture both short-term patterns and long-term dependencies. A 2D convolutional layer with 32 channels and kernel size 7 first processes the multivariate input to capture interactions between different features and temporal patterns. This is followed by a GRU layer with 64 hidden units that models long-term dependencies in the employment trends. To handle the cyclical nature of employment data\cite{burns1946measuring}, we implement skip connections at 4-step (weekly) and 24-step (monthly) intervals, complemented by a 7-day autoregressive component for modeling short-term linear trends.

For each industry sector, our processing pipeline follows these steps:
\begin{itemize}
    \item Normalizing input features and creating 30-day sliding window sequences
    \item Producing both a 7-day employment change forecast and an Industry Employment Health Index (IEHI)
    \item Assessing performance using MAPE and RMSE for forecasting accuracy, and Spearman's rank correlation for IEHI validation
\end{itemize}

For example, when processing the manufacturing sector, the model ingests a 30-day window of all input features, processes them through the convolutional and GRU layers to capture both immediate and long-term patterns, and outputs both the next week's employment forecast and an IEHI score. The IEHI is computed by combining the predicted employment stability with static factors like skill requirements and turnover rates. This dual-output approach provides both actionable short-term predictions and meaningful long-term industry health assessments.

The training process employs Mean Squared Error loss for the forecasting task, with regular validation against held-out data to ensure generalization. To account for industry-specific characteristics, we incorporate seasonal adjustment factors and industry-specific performance metrics in our evaluation framework.

\subsection{Architecture Overview}
\label{sec:architecture_overview}
Figure~\ref{fig:model_architecture} presents a high-level overview of our LSTNet-based architecture, which is designed to jointly forecast short-term employment change and assess long-term industry health.

The model accepts a multivariate input sequence $X \in \mathbb{R}^{T \times F}$, where $T = 30$ denotes the sliding window length and $F$ is the number of input features (e.g., employment rate, CPI, job openings). This input is represented in the architecture figure as \textbf{Item 1}.

The input is first processed by a 2D convolutional layer (\textbf{Item 2}) to extract local temporal patterns across variables, followed by a GRU layer with 64 hidden units (\textbf{Item 3}) to model long-term temporal dependencies. To account for periodic structure in labor dynamics, we introduce skip connections at weekly (4-step) and monthly (24-step) intervals (\textbf{Item 4}). 

A parallel autoregressive (AR) component $\hat{y}_t^{\text{AR}}$ (\textbf{Item 5}) enhances the model’s ability to capture short-term linear trends, complementing the GRU’s non-linear modeling.

The outputs of the GRU, skip connection module, and AR component are concatenated (\textbf{Item 6}) and passed through two heads: an employment forecast head $\hat{y}_t$ (\textbf{Item 7}) and a module that computes the Industry Employment Health Index (IEHI), denoted $s^{\text{IEHI}}$ (\textbf{Item 8}). The IEHI is derived from smoothed volatility in predictions and sector-specific metadata such as separation rates and hiring stability.

This architecture enables the model to capture both short-term and long-term dynamics in labor markets while producing interpretable metrics that inform sector-level workforce planning.

\begin{figure}[ht]
 \centering
 \includegraphics[width=1\linewidth]{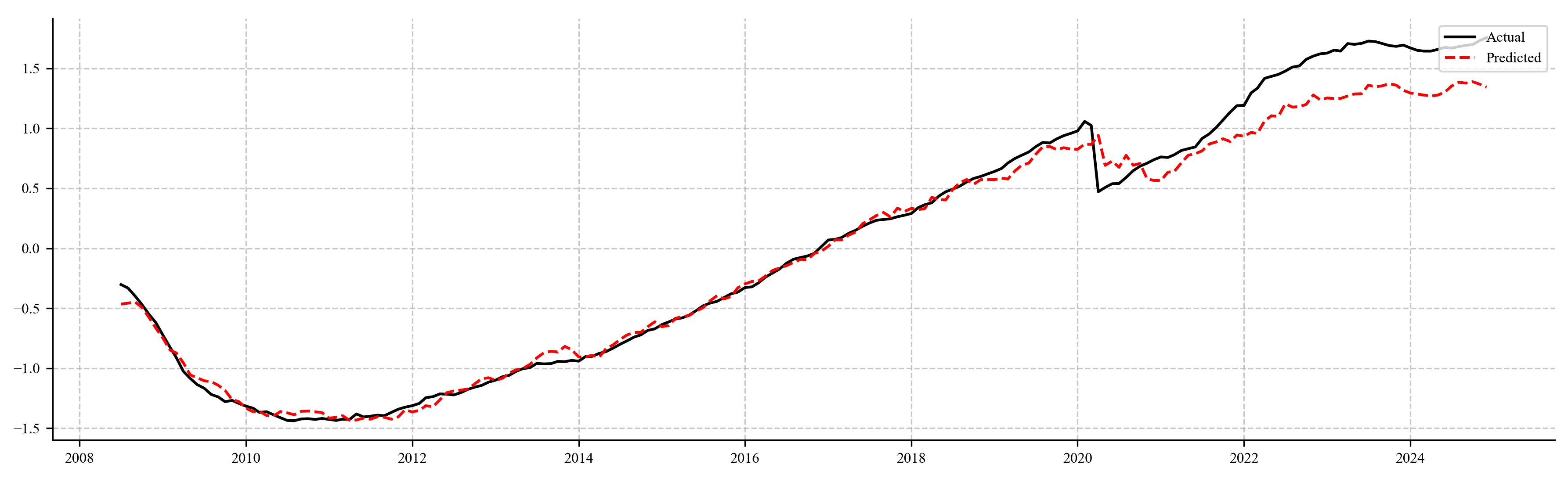}
 \caption{Employment in Financial Activities - Actual vs. Predicted.}
 \label{fig:financial_employment_pred}
\end{figure}

\begin{table}[htbp]
    \centering
    \caption{Model Performance by Industry on Employment Forecasting}
    \label{tab:industry_model_performance}
    \begin{tabular}{lccc}
        \toprule
        \textbf{Industry} & \textbf{MSE} & \textbf{MAE} & \textbf{SMAPE (\%)} \\
        \midrule
        Construction & 0.0454 & 0.1284 & 21.44 \\
        Education/Health & 0.0358 & 0.0980 & 13.08 \\
        Financial Activities & 0.0290 & 0.1117 & 13.42 \\
        Information & 0.1383 & 0.2051 & 41.09 \\
        Leisure/Hospitality & 0.2787 & 0.1400 & 18.13 \\
        Manufacturing & 0.0572 & 0.1443 & 51.32 \\
        Natural Resources & 0.1816 & 0.2793 & 52.94 \\
        Other Services & 0.4272 & 0.2331 & 33.72 \\
        Professional Services & 0.0189 & 0.0737 & 16.94 \\
        Transport/Utilities & 0.1653 & 0.2309 & 32.96 \\
        \bottomrule
    \end{tabular}
\end{table}

\section{Results and Analysis}
\label{sec:Results_Analysis}

Our system was evaluated on industry-specific employment data using key labor indicators and weekly metrics. The LSTNet model was trained on employment data from March 2008 to December 2024 and tested on recent weeks, producing both short-term forecasts and an Industry Employment Health Index (IEHI) for each sector.

To visually demonstrate some examples of model performance, \cref{fig:financial_employment_pred} showcases an example of an acceptable prediction model, while \cref{fig:natural_resources_pred} showcases rather poor performance. Most models fail to generalize around 2020, or when the Coronavirus pandemic caused mass unpredictability across all industries. This section will contain both specific industry comparisons and generalized model assessments to demonstrate the effectiveness of the model in various contexts.

\subsection{Forecast Performance Across Sectors}

The model demonstrated strong predictive accuracy in sectors with stable or seasonally structured employment trends. \textit{Education and Health Services} achieved the best overall performance, with a SMAPE below 13\%, driven by consistent post-recession recovery and low post-pandemic volatility. Similarly, \textit{Financial Activities} and \textit{Professional and Business Services} exhibited low error rates, with the model capturing both seasonal cycles and macroeconomic trends with minimal lag. These sectors also maintained the lowest mean absolute errors, as shown in Table~\ref{tab:industry_model_performance}. A full visual ranking of SMAPE values across all industries is provided in \cref{fig:model_performance}, where sectors are ordered from least to greatest error. 

In contrast, sectors with high structural volatility faced greater challenges. \textit{Information}, \textit{Manufacturing}, and \textit{Natural Resources and Mining} experienced an elevated forecasting error due to irregular post-COVID rebounds and idiosyncratic sectoral shocks. For example, in the Natural Resources sector (Figure~\ref{fig:natural_resources_pred}), the model under-predicts during sudden employment surges, reflecting difficulties in learning highly nonstationary patterns.

Sectors such as \textit{Leisure and Hospitality} and \textit{Construction}, which experienced deep pandemic-era contractions followed by uneven recoveries, yielded intermediate results. While the model effectively recovered alignment with post-2021 trends, it exhibited degraded performance during the discontinuous rebounds of 2020–2021. Sectors with mixed seasonal and structural characteristics—such as \textit{Transportation and Utilities} and \textit{Other Services}—were forecast with moderate accuracy. In these cases, the model tracked long-term growth, while occasionally lagging behind sharp inflections tied to short-term hiring or separation surges.

The model performed decently across all industries, with an average SMAPE of 26.7\%, as pictured in \cref{fig:All Industries Employment Over Time}. In this case, the model's predictions across all industries were averaged, and used to make predictions about the general market. This is essentially a reduction in data quality, so the elevated SMAPE reflects the loss of sector-specific structure and volatility patterns that the model relies on for precision.

\subsection{IEHI Alignment and Rank Correlation}

To evaluate the consistency between predictive accuracy and our interpretability metric, we computed Spearman's rank correlation coefficient ($\rho$) between industry-level SMAPE scores and IEHI rankings. The IEHI was constructed by aggregating stability and volatility signals derived from the predicted employment trajectories of each sector (see Section~\ref{sec:mainApproach}). Industries with low volatility, sustained growth, and minimal turnover were ranked higher. Spearman’s $\rho$ was used to quantify the degree of ordinal agreement between IEHI rank and actual model performance. As shown in Table~\ref{tab:combined_iehi_spearman}, we observe a strong and statistically significant correlation ($\rho = 0.95$, $p = 2.28 \times 10^{-5}$), indicating that the IEHI effectively captures relative forecasting difficulty and may serve as a meaningful proxy for employment health and predictability.

\subsection{Interpretation of Results}
The LSTNet model demonstrated strong predictive performance in sectors characterized by stable employment trajectories and regular seasonal patterns. Notably, \textit{Education and Health Services}, \textit{Financial Activities}, and \textit{Professional and Business Services} yielded the lowest SMAPE scores (13.1\%, 13.4\%, and 16.9\%, respectively), reflecting the model’s ability to capture gradual growth and cyclical behavior with high fidelity.

In contrast, sectors with high structural variability—such as \textit{Natural Resources and Mining} (52.9\%), \textit{Manufacturing} (51.3\%), and \textit{Information} (41.1\%)—posed greater forecasting challenges. These industries experienced abrupt shifts driven by macroeconomic shocks\cite{sahin2014mismatch}, supply chain disruptions, or technological transformation, which degraded point forecast accuracy. Nonetheless, even in these volatile contexts, the model generally preserved directional trends and demonstrated reliable ordinal ranking of sectoral employment changes.

While average model SMAPE across industries was 26.7\%, this figure was heavily influenced by outlier sectors with extreme volatility. Outside of rare periods of economic turbulence—such as the 2020 downturn—model predictions aligned well with observed employment patterns over both medium and long-range horizons.

Finally, the LSTNet model consistently outperformed the persistence baseline across all industries and closely approached oracle-level performance in low-volatility domains. These results affirm the model’s ability to synthesize multivariate labor signals—such as job openings, separations, and average hours worked—into accurate and interpretable forecasts of sector-specific employment dynamics.

\begin{figure}[ht]
 \centering
 \includegraphics[width=1\linewidth]{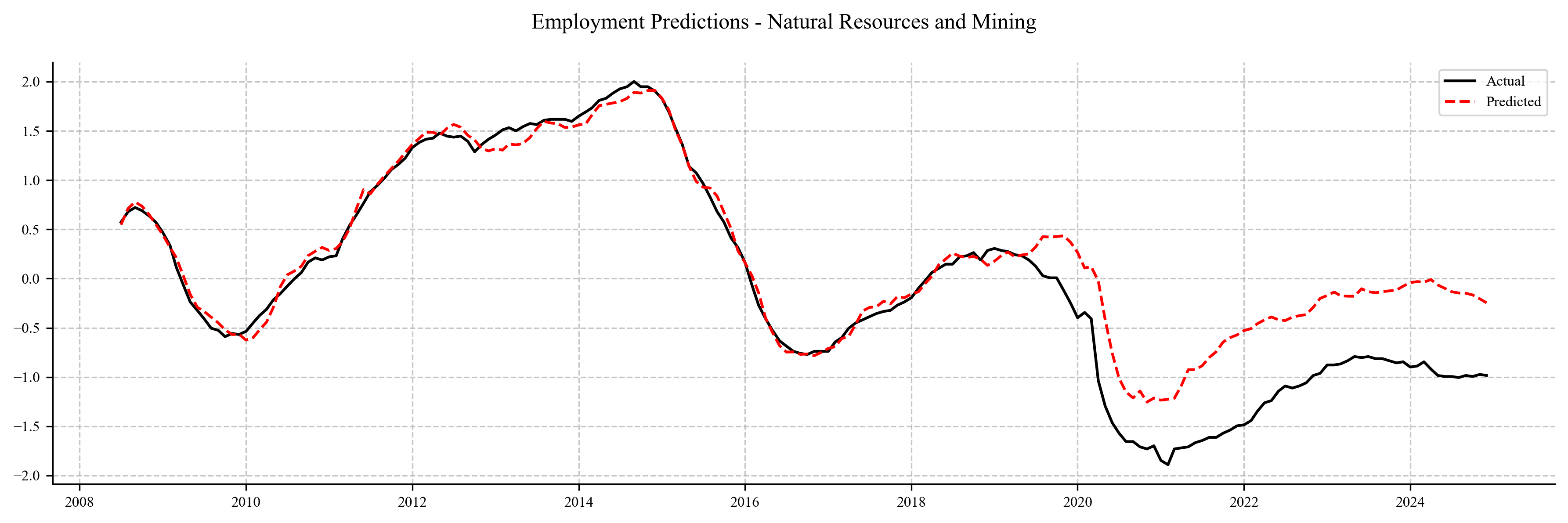}
 \caption{Employment in Natural Resources and Mining - Actual vs. Predicted.}
 \label{fig:natural_resources_pred}
\end{figure}

\begin{table}[htbp]
    \centering
    \caption{SMAPE and IEHI Rank by Industry with Spearman Correlation}
    \label{tab:combined_iehi_spearman}
    \begin{tabular}{lcc}
        \toprule
        Industry & SMAPE (\%) & IEHI Rank \\
        \midrule
        Financial Activities & 13.42 & 1 \\
        Education Services & 13.08 & 2 \\
        Professional Services & 16.94 & 3 \\
        Construction & 21.44 & 4 \\
        Transportation & 32.96 & 5 \\
        Leisure/Hospitality & 18.13 & 6 \\
        Other Services & 33.72 & 7 \\
        Information & 41.09 & 8 \\
        Manufacturing & 51.32 & 9 \\
        Natural Resources & 52.94 & 10 \\
        \midrule
        \multicolumn{3}{l}{\textit{Spearman Rank Correlation}} \\
        \cmidrule(lr){1-2}
        Spearman's $\rho$ & 0.95 & \\
        $p$-value & 2.28e-05 & \\
        \bottomrule
    \end{tabular}
\end{table}

\subsection{Error Analysis}

To better understand the model's failure modes and temporal behavior, we conducted a detailed error analysis across time periods and industry types. A consistent observation across sectors was the model’s reduced performance during periods of extreme volatility, particularly surrounding macroeconomic disruptions. For example, in \textit{Natural Resources and Mining}, large post-2020 employment swings led to substantial over- and under-predictions, as shown in Figure~\ref{fig:natural_resources_pred}. These errors reflect the model's limited ability to extrapolate beyond historical precedent, especially during structural regime shifts.

Temporal lag was also observed in the model’s responsiveness to inflection points. In sectors like \textit{Transportation and Utilities}, sharp employment reversals were often predicted with a delay of one to two months. While skip connections and autoregressive components in LSTNet are designed to capture local and periodic patterns, these mechanisms proved insufficient for capturing sudden trend shifts in real-time. This delay highlights a potential need for leading indicator integration or attention-based mechanisms to reduce temporal inertia.

Another class of errors stemmed from signal degradation in data-sparse or structurally noisy industries. In \textit{Information} and \textit{Other Services}, high-frequency volatility and inconsistent recovery trajectories degraded SMAPE performance despite the model's correct capture of general direction. These findings suggest that the model performs best when data is both seasonally structured and statistically stationary.

\section{Future Work}
\label{sec:Description_Challenges}

While the current approach provides a strong baseline for industry-specific employment forecasting, several directions remain open for further exploration. One avenue involves incorporating causal inference methods to better capture the impact\cite{brodersen2015inferring} of external disruptions, such as policy shifts or macroeconomic shocks, which are often missed by conventional sequence models. Furthermore, leveraging spatio-temporal architectures could improve generalization across regions by modeling the geographic diffusion of labor trends.

Improving interpretability is another focus. We aim to develop tools that reveal the most influential features driving forecast changes, creating a more trustworthy and transparent model. In parallel, our goal is to incorporate uncertainty quantification, which would give a more accurate representation of the strength of the identified trend.

We are exploring extensions to the Industry Employment Health Index, such as hierarchical designs that blend forecasts across different temporal resolutions (e.g., weekly, monthly, quarterly). This could yield a more nuanced and responsive index that better reflects the layered structure of labor dynamics.
\section*{Author Contributions}

Adam Nelson-Archer developed the models, implemented the baseline and oracle evaluations, and outlined the paper. Aleia Sen prepared the dataset and data integrations. Meena Al Hasani contributed to results analysis. Sofia Davila and Jessica Le assisted with writing the model and analysis sections. Omar Abbouchi wrote the Future Work section. All authors reviewed and edited the final manuscript.

% \small
\bibliographystyle{ieeenat_fullname}
\bibliography{main}

 \clearpage
\setcounter{page}{1}

\maketitlesupplementary

\section{Appendix}
\label{sec:appendix}

The following appendix provides additional material that complements the main body of the paper. For clarity, only select visualizations and summaries were presented earlier.

\subsection{Code and Github Repository}
The code used for the creation of our model, the evaluation of our results, and the processing of our database can be found at this GitHub repository: \url{https://github.com/aleia-s/4337_proj}. 

Our model training and evaluation files can be found inside of the \texttt{src} folder, while our databases and related code can be found in the \texttt{data} and \texttt{notebooks} folders. The models used in this paper can be found in the \texttt{models} folder.

\subsection{Model Hyperparameters}

We include the full configuration used to train and evaluate the LSTNet model. These settings reflect architectural choices, training parameters, data paths, and visualization options used in our pipeline. This inclusion serves as a reproducibility aid for future work and reviewers.

\begin{lstlisting}[language=Python, caption={Configuration settings for the LSTNet model and training process}]
# Model parameters
MODEL_CONFIG = {
    'cnn_kernel_size': 6,  
    'rnn_hidden_size': 100, 
    'skip_size': 24, 
    'skip_hidden_size': 5,
    'highway_window': 24, 
}

TRAINING_CONFIG = {
    'batch_size': 128,
    'epochs': 100,
    'learning_rate': 0.001,
    'sequence_length': 28, 
    'test_size': 0.2,
    'val_size': 0.2,
}

\end{lstlisting}

\subsection{Extended Forecast Visualizations}

\begin{figure}[ht]
 \centering
 \includegraphics[width=1\linewidth]{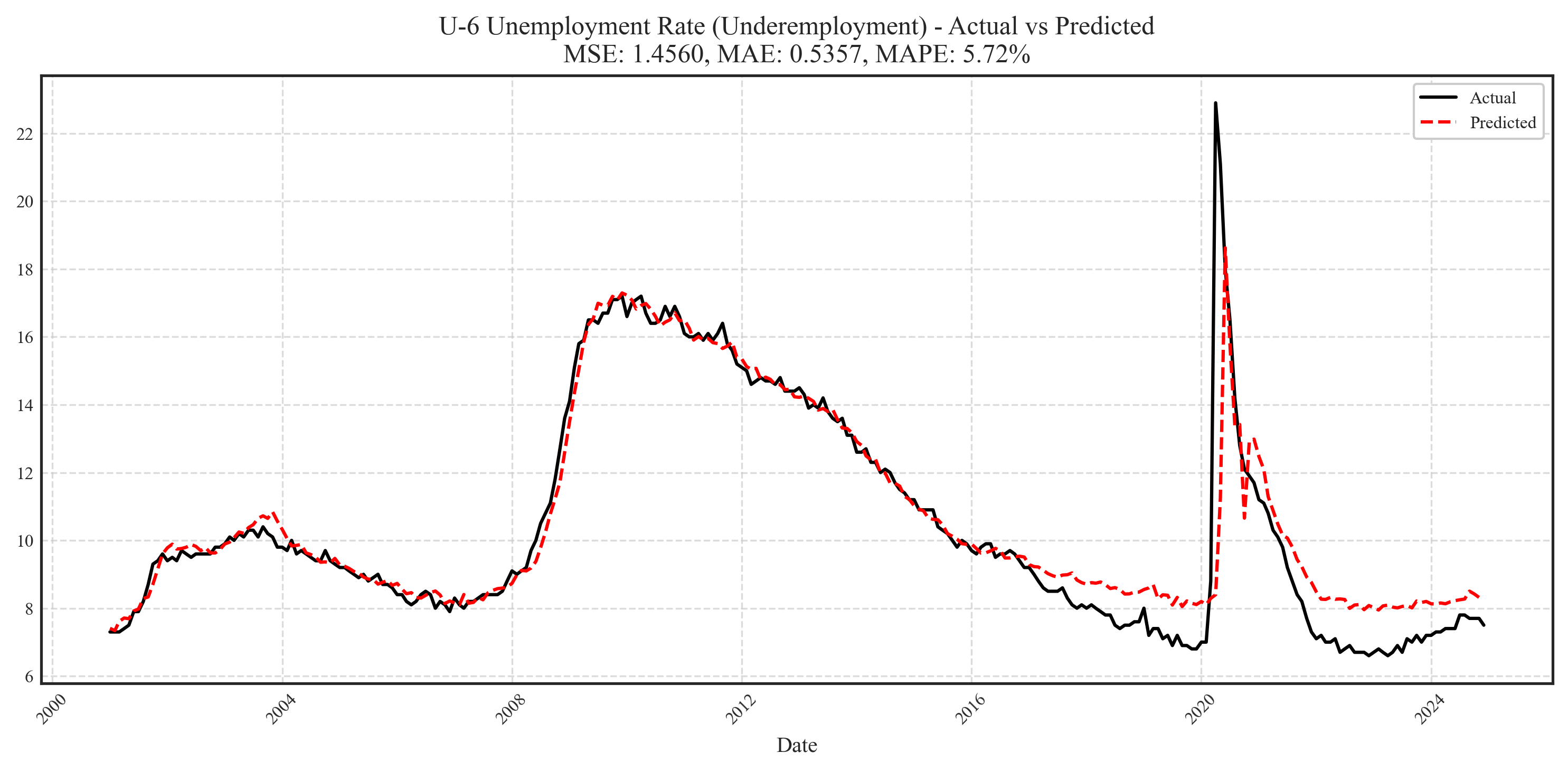}
 \caption{U6 (Underemployment) - Actual vs. Predicted.}
 \label{fig:u6_graph}
\end{figure}

\begin{figure}[ht]
 \centering
 \includegraphics[width=1\linewidth]{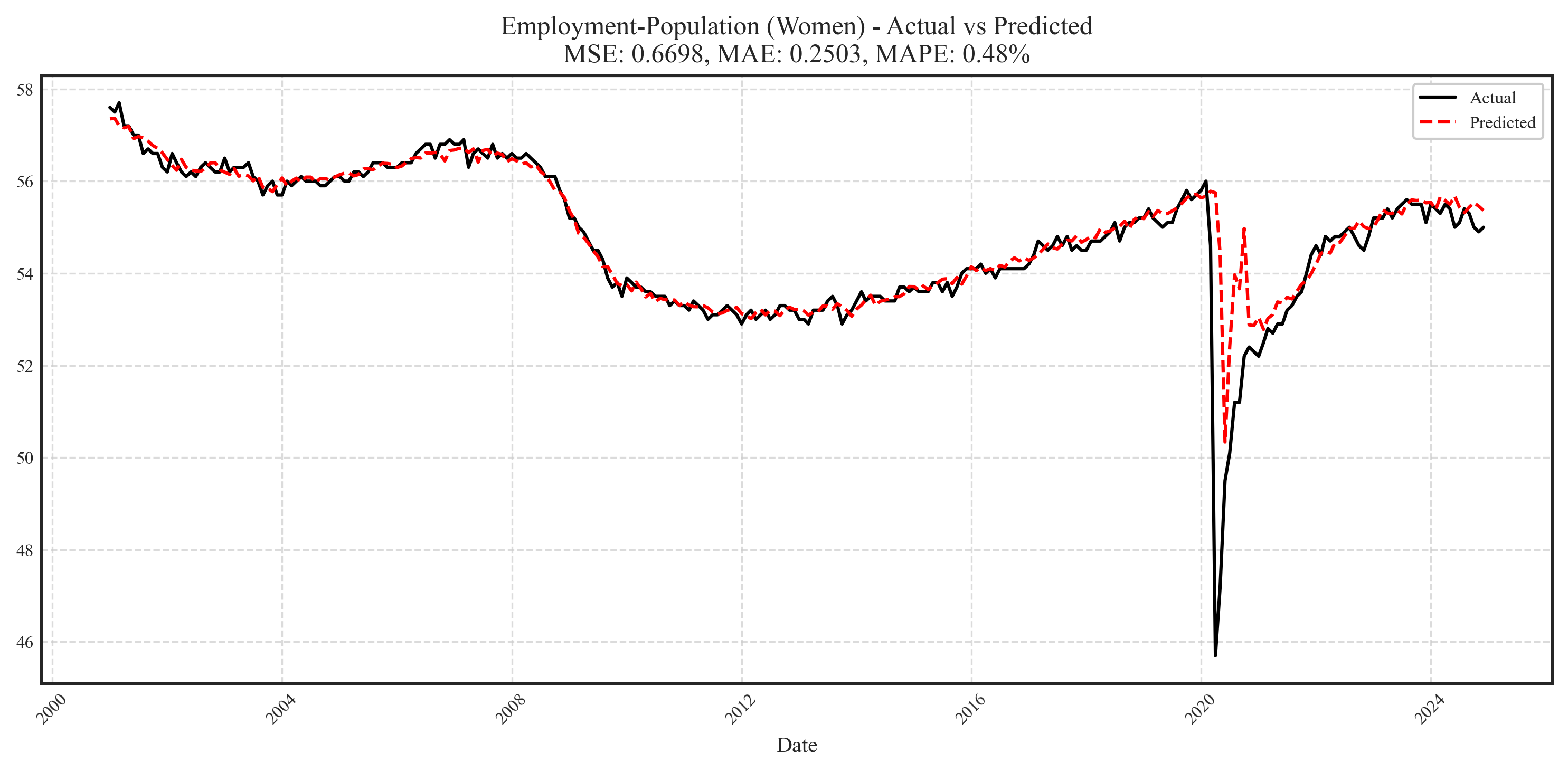}
 \caption{Unemployment Rate (Women) - Actual vs. Predicted.}
 \label{fig:unemp_women}
\end{figure}

\begin{figure}[ht]
 \centering
 \includegraphics[width=1\linewidth]{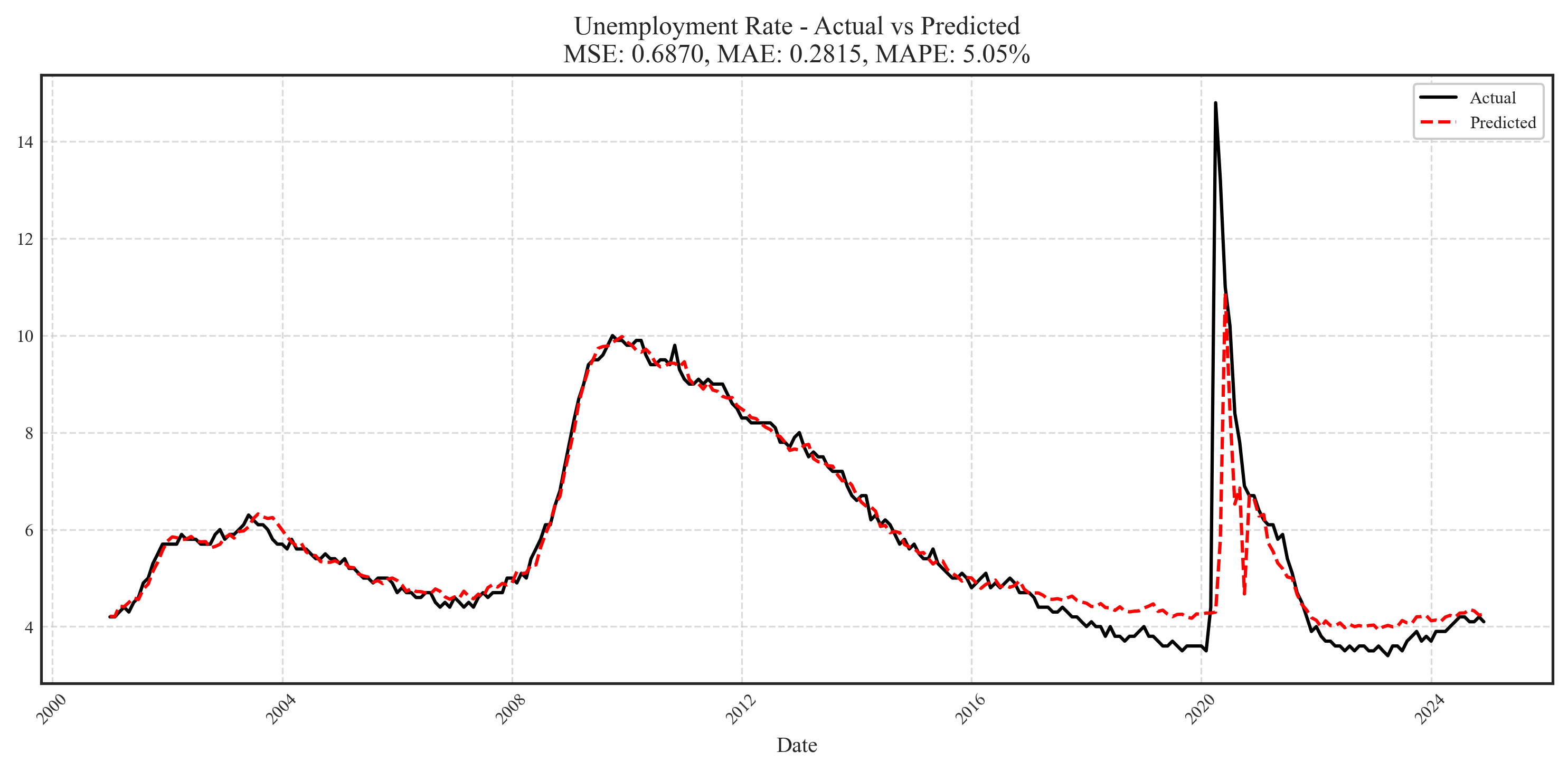}
 \caption{Unemployment Rate (Overall) - Actual vs. Predicted.}
 \label{fig:unemp_rate}
\end{figure}

The main body of this report focuses on the most representative results from our forecasting model, but due to space constraints, not all outputs could be included. In this appendix, we present full-length visualizations of additional employment indicators whose abbreviated forms appeared in Section~\ref{sec:Results_Analysis}. These expanded plots demonstrate the model’s performance across a range of demographic and labor-related metrics, providing further insight into its strengths and weaknesses. Some of these plots (such as \cref{fig:u6_graph}, \cref{fig:unemp_women}, and \cref{fig:unemp_rate}) go beyond the scope of our paper, and evaluate the generalization of the forecasting model to unemployment indicators, included as additional model stress tests.

\begin{figure}[ht]
 \centering
 \includegraphics[width=1\linewidth]{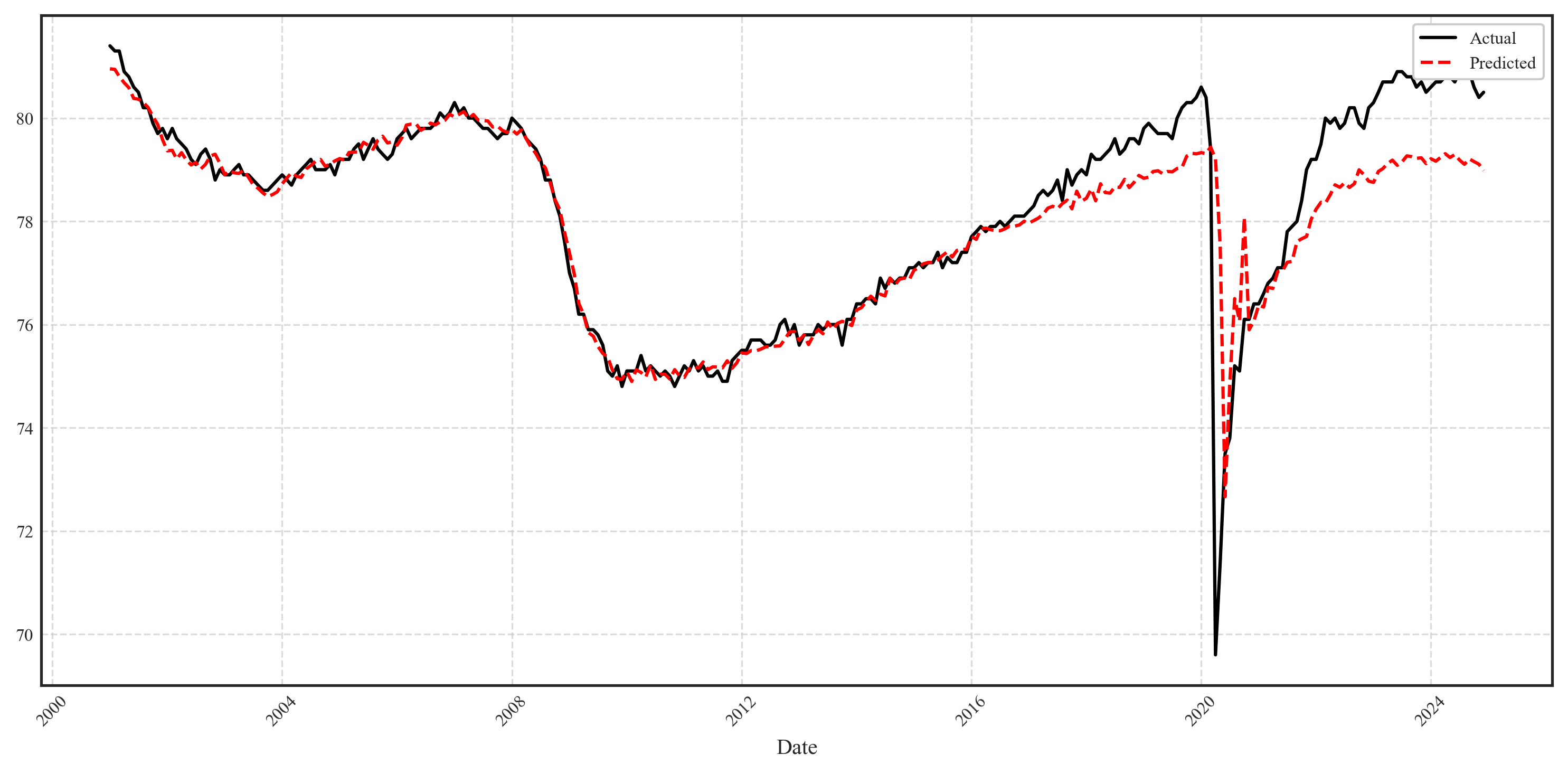}
 \caption{Unemployment Rate (Men) - Actual vs. Predicted.}
 \label{fig:unemp_men}
\end{figure}

\begin{figure}[ht]
 \centering
 \includegraphics[width=1\linewidth]{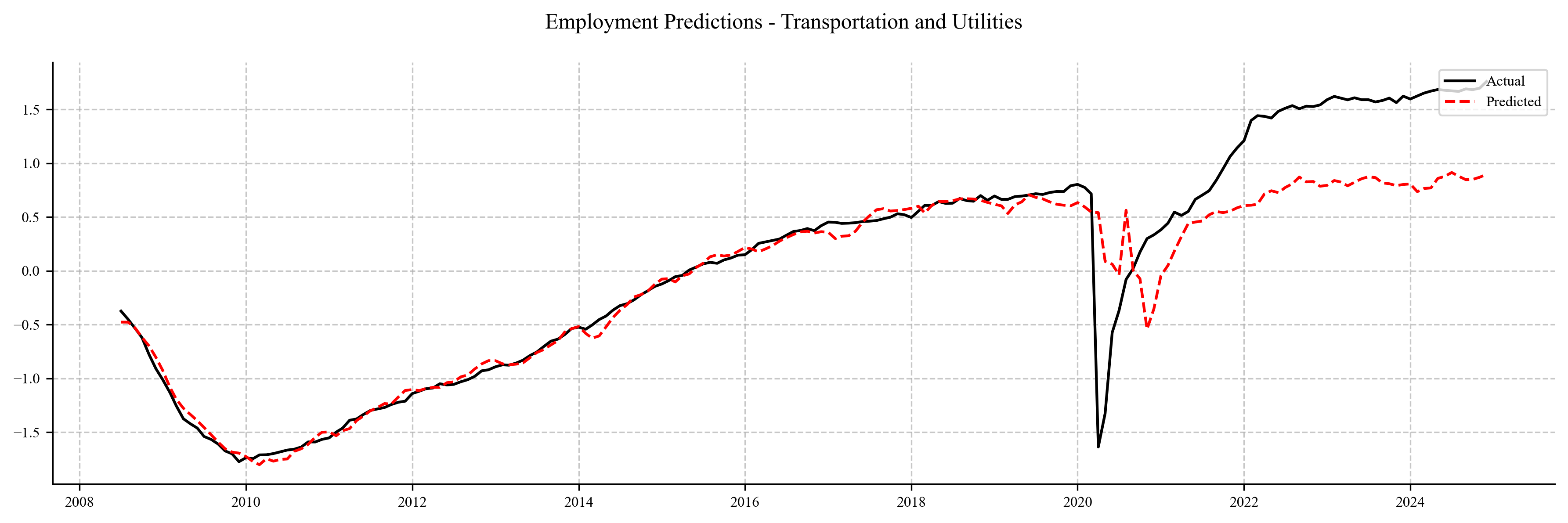}
 \caption{Changes in Employment in the Utilities Industry - Actual vs. Predicted.}
 \label{fig:utilities}
\end{figure}

\begin{figure}[ht]
 \centering
 \includegraphics[width=1\linewidth]{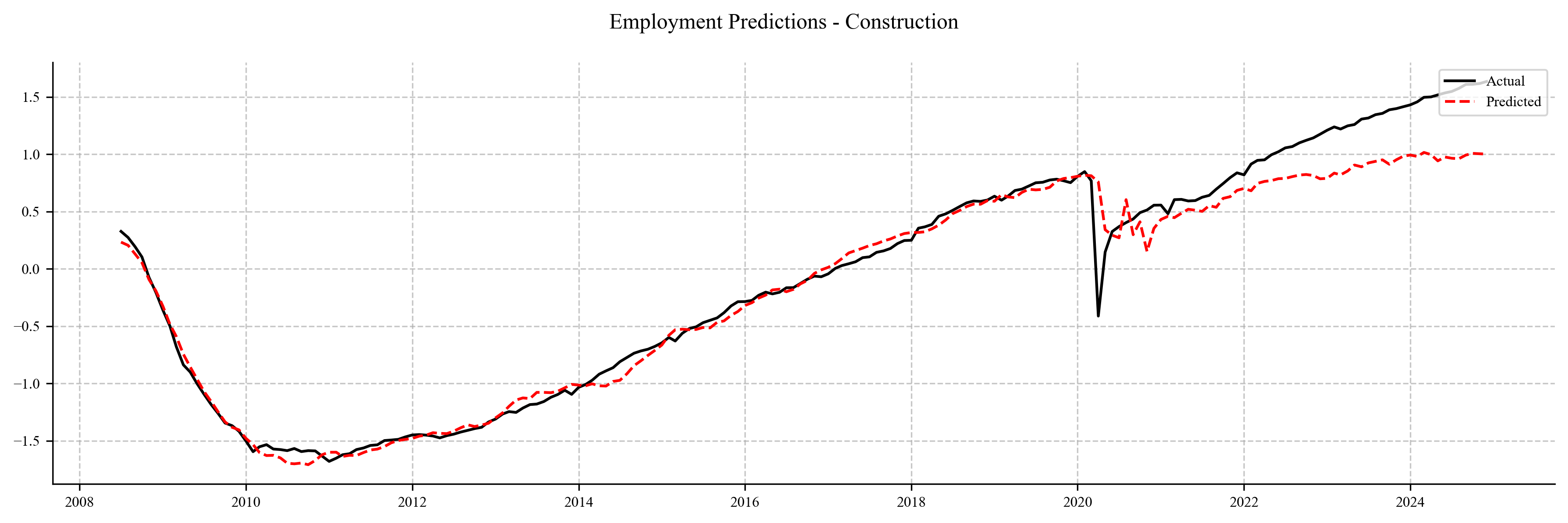}
 \caption{Changes in Employment in the Construction Industry - Actual vs. Predicted.}
 \label{fig:construction}
\end{figure}

\begin{figure}[ht]
 \centering
 \includegraphics[width=1\linewidth]{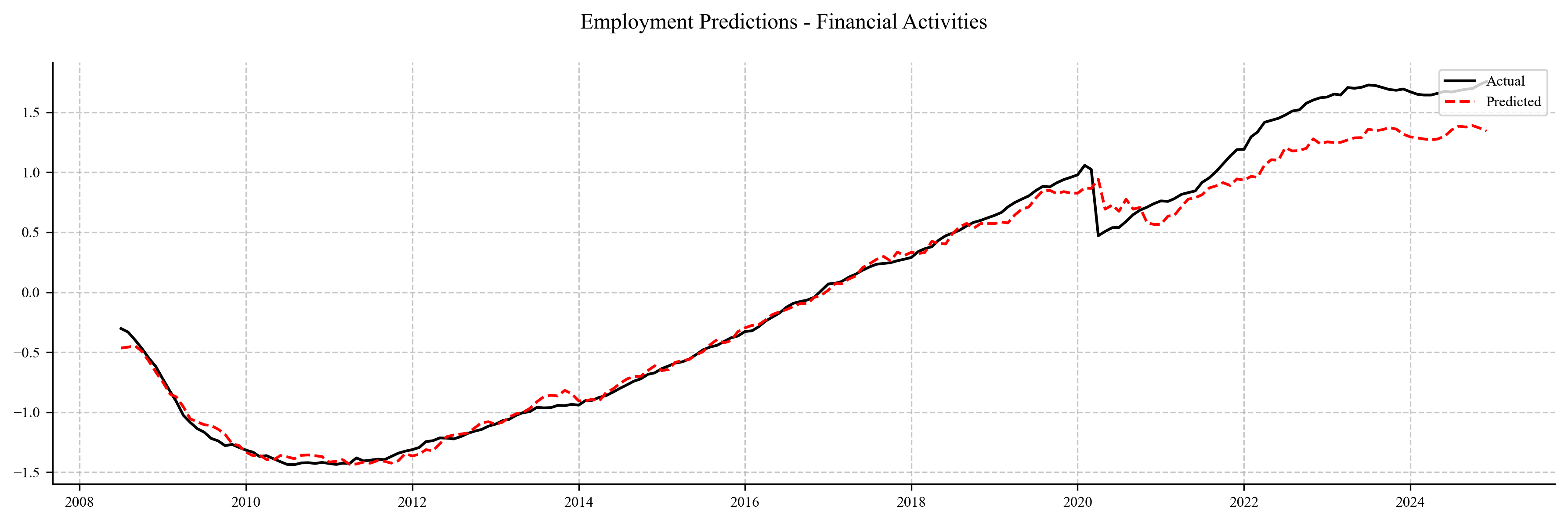}
 \caption{Changes in Employment in the Financial Industry - Actual vs. Predicted.}
 \label{fig:financial}
\end{figure}

\begin{figure}[ht]
 \centering
 \includegraphics[width=1\linewidth]{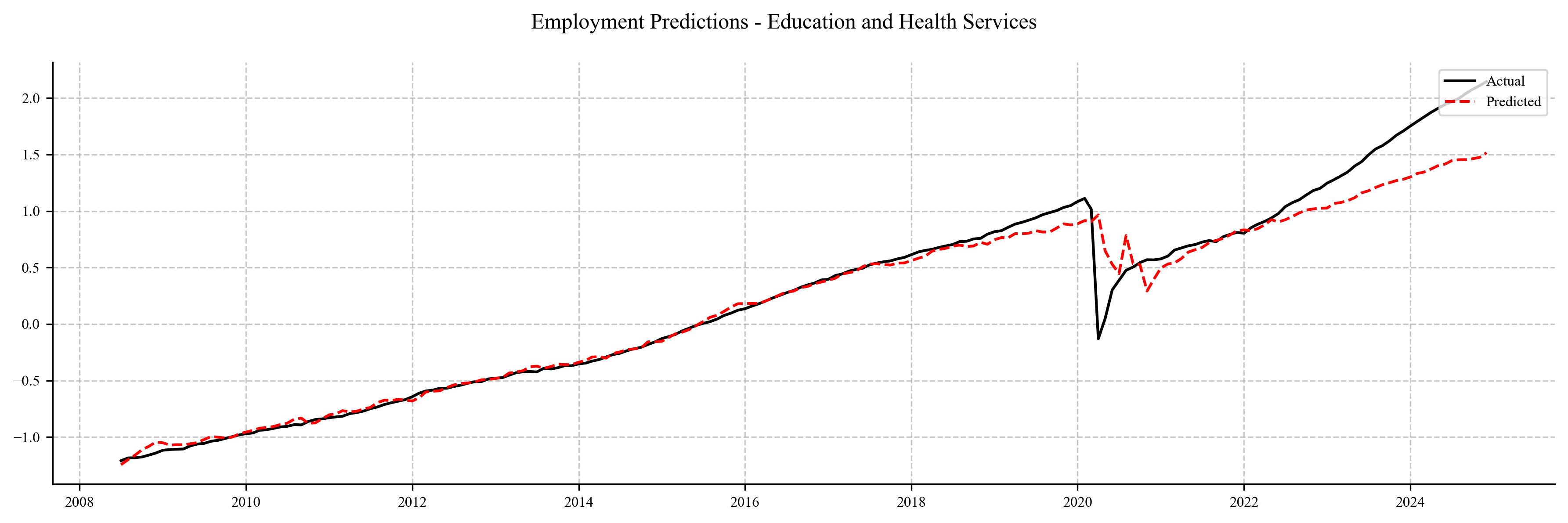}
 \caption{Changes in Employment in the Education Industry - Actual vs. Predicted.}
 \label{fig:education}
\end{figure}

\subsection{Industry-Level Trend Analysis}

To supplement our forecasting results, we conducted exploratory data analysis on industry-level employment indicators. These visualizations provide insight into sector-specific, inter-industry relationships, and volatility in employment behavior. 

\begin{figure}
    \centering
    \includegraphics[width=1\linewidth
]{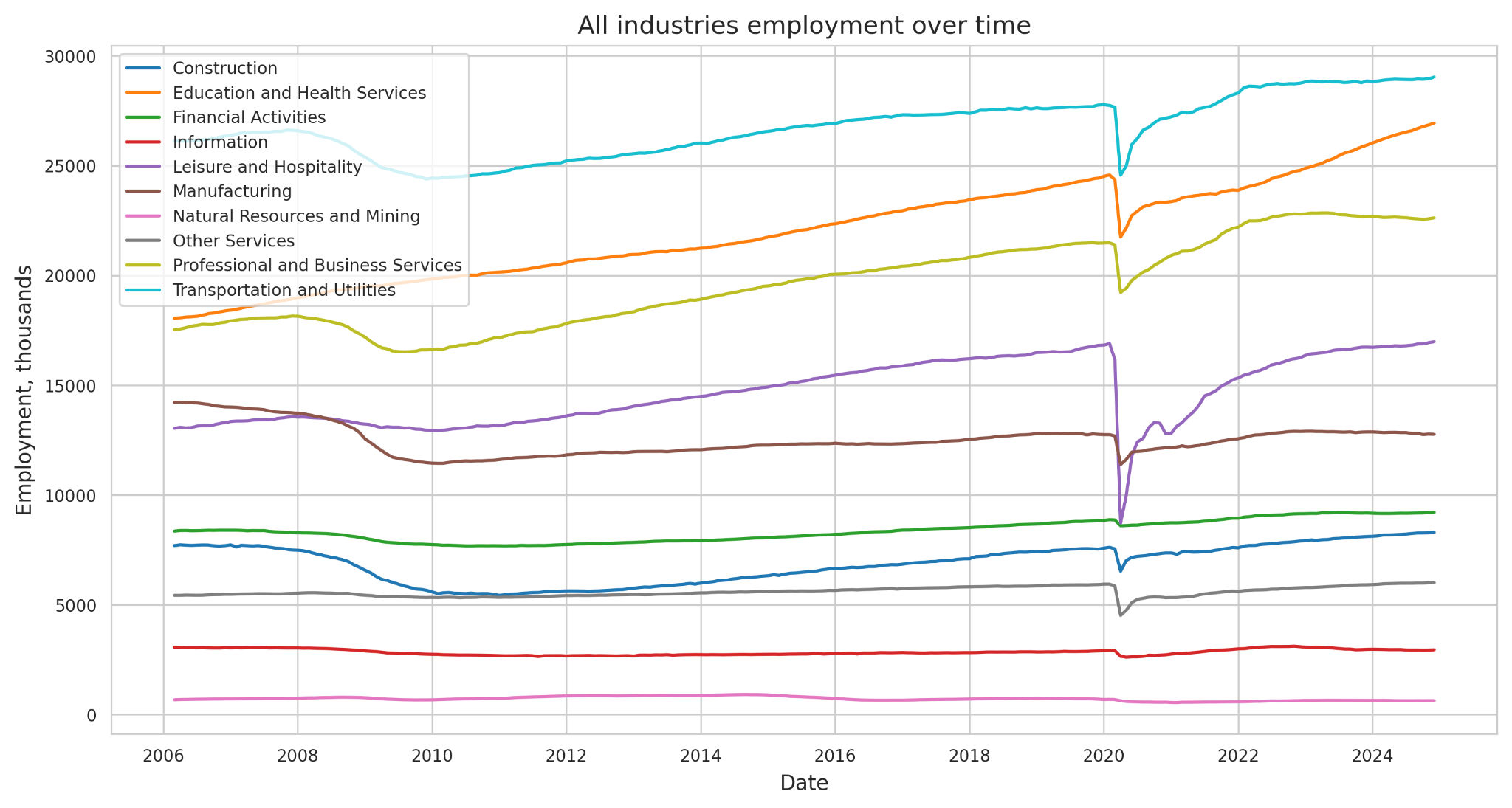}
    \caption{All Industries Employment Over Time}
    \label{fig:All Industries Employment Over Time}
\end{figure}

Fig. \ref{fig:All Industries Employment Over Time} shows employment levels for 11 major industries. While most sectors exhibit long-term growth, sharp drops in 2020 highlight differential recovery trajectories. For instance, Leisure and Hospitality experienced a steep decline followed by a strong rebound, while Manufacturing showed a slower recovery.
 
\begin{figure}
    \centering
    \includegraphics[width=1\linewidth]{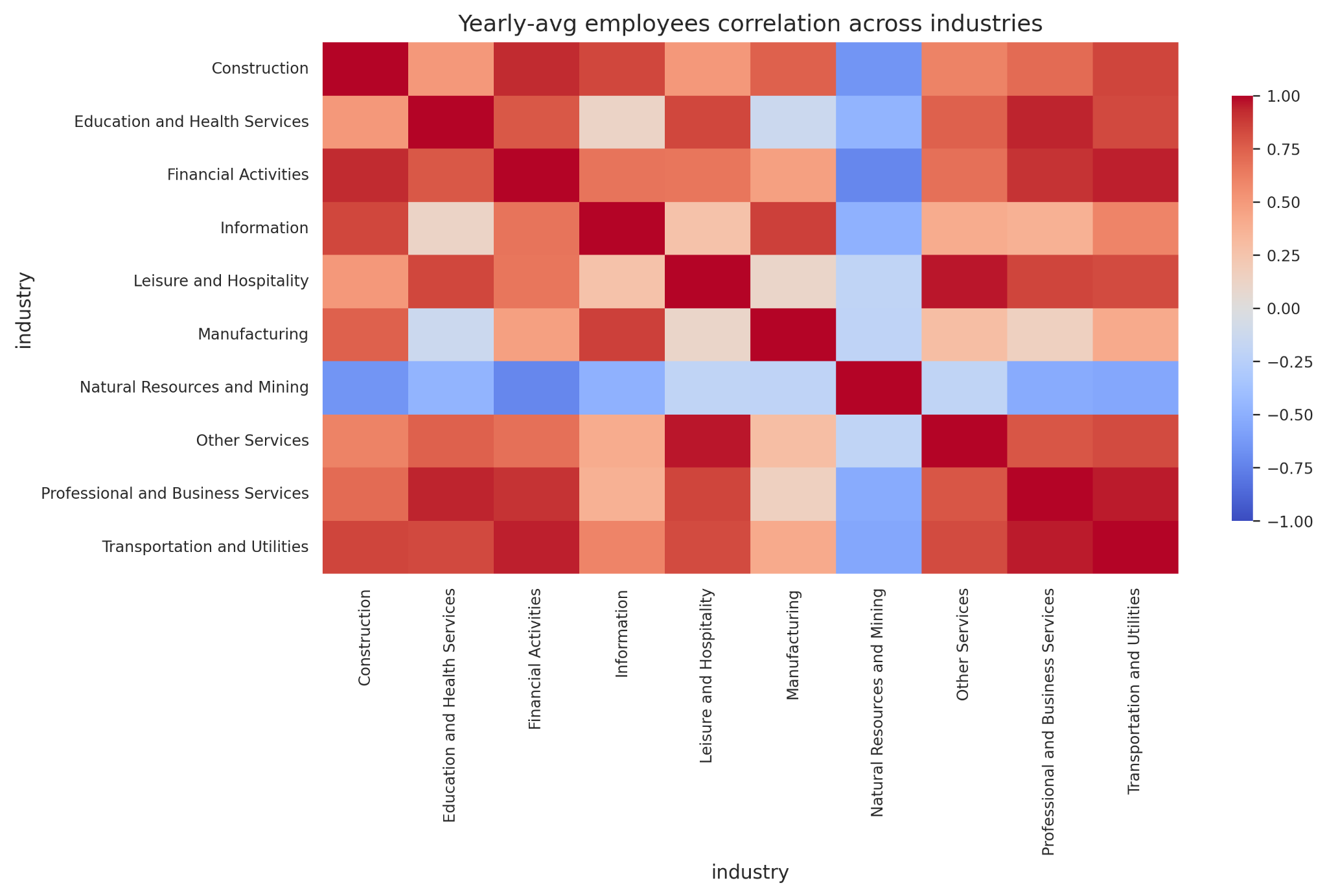}
    \caption{Yearly-Average Correlation Matrix Across Industries}
    \label{fig:Yearly-Average Correlation Matrix Across Industries}
\end{figure}

Fig. \ref{fig:Yearly-Average Correlation Matrix Across Industries} displays correlation coefficients between industry employment levels. Strong positive correlations were observed between service-oriented sectors such as Education \& Health Services and Transportation. In contrast, industries like Natural Resources and Manufacturing had weaker correlations with others, supporting our motivation for industry-specific modeling or clustering.
 
\begin{figure}
    \centering
    \includegraphics[width=1\linewidth]{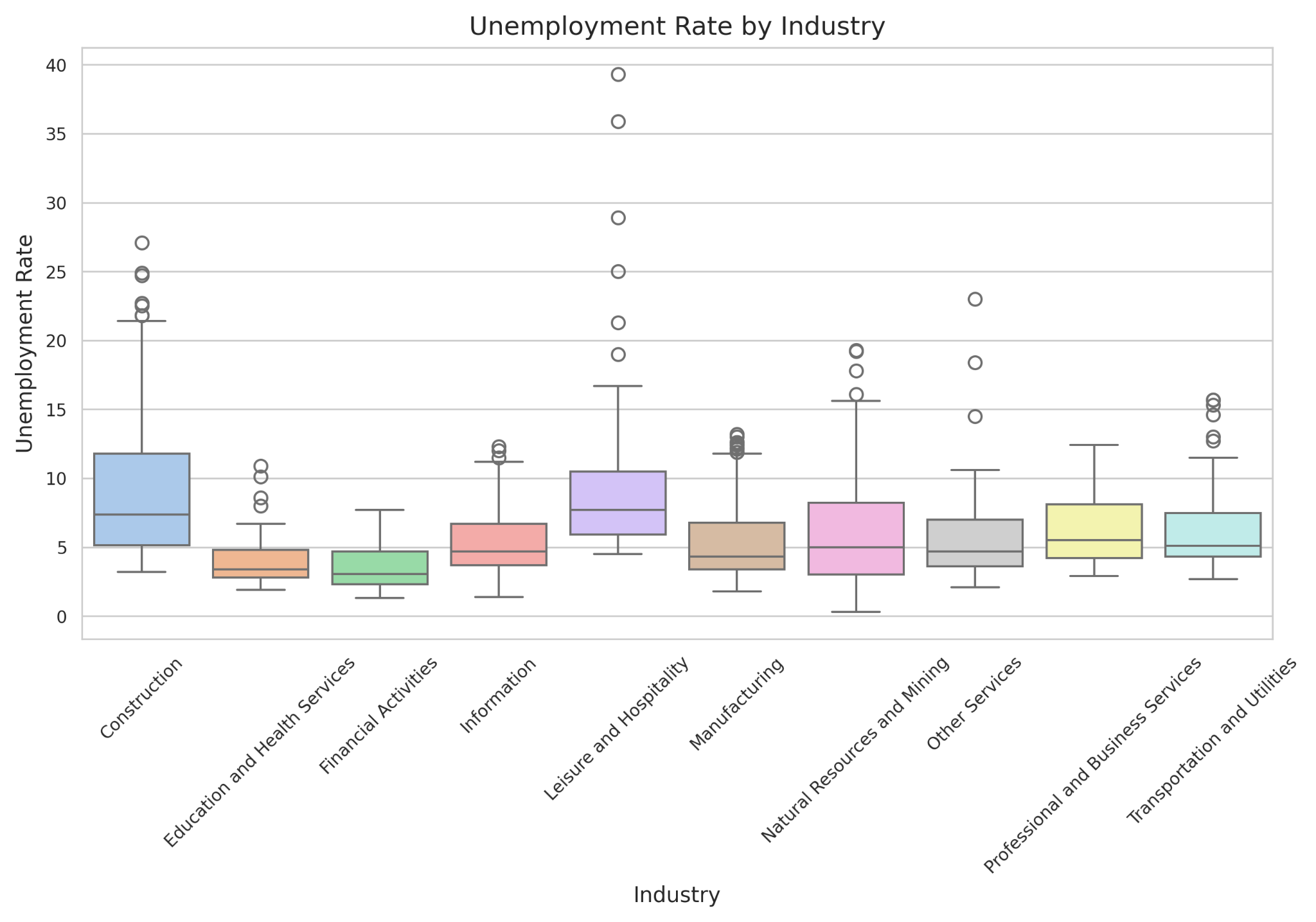}
    \caption{Unemployment Rate by Industry}
    \label{fig:Unemployment Rate by Industry}
\end{figure}

Fig. \ref{fig:Unemployment Rate by Industry} captures unemployment rate distributions by industry. Sectors like Leisure \& Hospitality and Construction showed higher variability and more frequent outliers, indicating higher structural risk and turnover. Conversely, Education and Financial Activities were among the most stable.

\end{document}